\documentclass[aps,prl,twocolumn,superscriptaddress]{revtex4-1}

\usepackage{mhchem,graphicx,longtable,xfrac}
\usepackage[colorlinks=true,citecolor=blue,linkcolor=blue,breaklinks=true]{hyperref}
\usepackage{amssymb}
\usepackage{graphicx}
\usepackage{amsmath}

\usepackage{times}
\usepackage{color}
\usepackage{bm}

\begin{document}

\newcommand {\beq} {\begin{equation}}
\newcommand {\eeq} {\end{equation}}
\newcommand {\bqa} {\begin{eqnarray}}
\newcommand {\eqa} {\end{eqnarray}}
\newcommand {\ba} {\ensuremath{b^\dagger}}
\newcommand {\Ma} {\ensuremath{M^\dagger}}
\newcommand {\psia} {\ensuremath{\psi^\dagger}}
\newcommand {\psita} {\ensuremath{\tilde{\psi}^\dagger}}
\newcommand{\lp} {\ensuremath{{\lambda '}}}
\newcommand{\A} {\ensuremath{{\bf A}}}
\newcommand{\Q} {\ensuremath{{\bf Q}}}
\newcommand{\kk} {\ensuremath{{\bf k}}}
\newcommand{\qq} {\ensuremath{{\bf q}}}
\newcommand{\kp} {\ensuremath{{\bf k'}}}
\newcommand{\rr} {\ensuremath{{\bf r}}}
\newcommand{\rp} {\ensuremath{{\bf r'}}}
\newcommand {\ep} {\ensuremath{\epsilon}}
\newcommand{\nbr} {\ensuremath{\langle ij \rangle}}
\newcommand {\no} {\nonumber}
\newcommand{\up} {\ensuremath{\uparrow}}
\newcommand{\dn} {\ensuremath{\downarrow}}
\newcommand{\rcol} {\textcolor{red}}

\newcommand{\nyo}{\ce{NaYbO_2}\ }
\newcommand{\jeff}{$J_{eff}=1/2$\ }
\newcommand{\uud}{$up$-$up$-$down$\ }


\title{Spin excitations in the frustrated triangular lattice antiferromagnet NaYbO$_2$}



\author{Mitchell M. Bordelon}
\affiliation{Materials Department, University of California, Santa Barbara, California 93106, USA}

\author{Chunxiao Liu}
\affiliation{Department of Physics, University of California, Santa Barbara, California 93106, USA}

\author{Lorenzo Posthuma}
\affiliation{Materials Department, University of California, Santa Barbara, California 93106, USA}

\author{P. M. Sarte}
\affiliation{California Nanosystems Institute, University of California, Santa Barbara, California 93106, USA}

\author{N. P. Butch}
\affiliation{NIST Center for Neutron Research, National Institute of Standards and Technology, Gaithersburg, Maryland 20899, USA}

\author{Daniel M. Pajerowski}
\affiliation{Neutron Scattering Division, Oak Ridge National Laboratory, Oak Ridge, TN 37831, USA}

\author{Arnab Banerjee}
\affiliation{Neutron Scattering Division, Oak Ridge National Laboratory, Oak Ridge, TN 37831, USA}

\author{Leon Balents}
\affiliation{Kavli Institute for Theoretical Physics, University of California, Santa Barbara, Santa Barbara, California 93106, USA}

\author{Stephen D. Wilson}
\email[]{stephendwilson@ucsb.edu}
\affiliation{Materials Department, University of California, Santa Barbara, California 93106, USA}


\date{\today}

\begin{abstract}
 Here we present a neutron scattering-based study of magnetic excitations and magnetic order in \nyo under the application of an external magnetic field. The crystal electric field-split $J=7/2$ multiplet structure is determined, revealing a mixed $|m_z \rangle$ ground state doublet and is consistent with a recent report Ding et al. \cite{ding_tsirlin}. Our measurements further suggest signatures of exchange effects in the crystal field spectrum, manifested by a small splitting in energy of the transition into the first excited doublet. The field-dependence of the low-energy magnetic excitations across the transition from the quantum disordered ground state into the fluctuation-driven ordered regime is analyzed. Signs of a first-order phase transition into a noncollinear ordered state are revealed at the upper-field phase boundary of the ordered regime, and higher order magnon scattering, suggestive of strong magnon-magnon interactions, is resolved within the previously reported \uud phase.  Our results reveal a complex phase diagram of field-induced order and spin excitations within NaYbO$_2$ and demonstrate the dominant role of quantum fluctuations cross a broad range of fields within its interlayer frustrated triangular lattice. 
\end{abstract}

\pacs{}

\maketitle

\section{I. Introduction}

\begin{figure*}[t]
	\includegraphics[width=\textwidth]{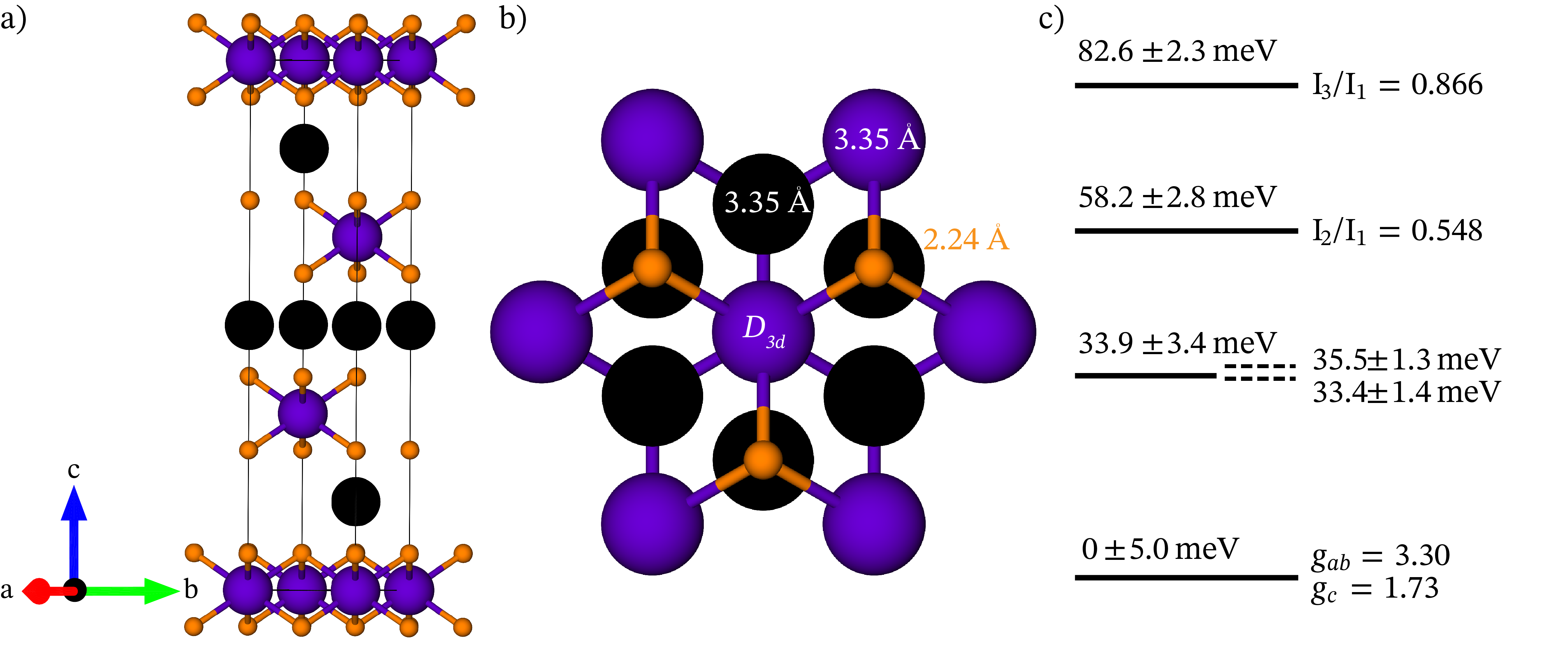}
	\caption{(a) The NaYbO$_2$ structure ($R\bar{3}m$, 1.6K \cite{paper1NYO}) contains alternating layers of equilateral triangular YbO$_6$ octahedra (Yb: purple, O: orange) and Na ions (black). (b) Top-down view of the local Yb$^{3+}$ $D_{3d}$ environment extended to two coordination shells. From the central Yb ion, six O$^{2-}$, six Na$^{+}$, and six Yb$^{3+}$ ions reside at 2.24, 3.35, and 3.35 \AA, respectively. (c) In the D$_{3d}$ environment, Yb creates four Kramers doublets as observed from Crystalline Electric Field (CEF) fits to inelastic neutron scattering data. Listed errors correspond to instrumental resolutions at $E_i$ = 150 meV (solid lines) and $E_i$ = 60 meV (dashed lines). The ground state has anisotropic $g$-factors of $g_c$ = 1.73 and $g_{ab}$ = 3.30 \cite{paper1NYO}. Dashed lines next to the first excited CEF doublet represent experimentally-observed splitting due to exhange-induced CEF dispersions observed in $E_i$ = 60 meV data. Intensity ratios of the second and third excitations (I$_2/$I$_1$, I$_3/$I$_1$) were obtained relative to the first CEF state. Uncertainties specified next to energy values are the calculated instrument energy resolution at those energies.}
	\label{fig:F1}
\end{figure*}

The triangular lattice antiferromagnet is a long-studied archetype of geometrically-driven magnetic frustration. It has been widely explored both experimentally and theoretically with goals to realize unconventional ground states that may arise from tailoring the degree of frustration, underlying anisotropies, and quantum fluctuations inherent to the moments decorating the magnetic lattice \cite{anderson1, anderson2, balents4, balents5, lee, witczak6, zhou7, cava_broholm}. While in the ideal Heisenberg limit, both classical and quantum moments develop three-sublattice 120$^{\circ}$ order \cite{j1j2_1, j1j2_2}, perturbing away from this limit realizes a rich phase space.  In particular, the search for conditions that realize ground states disordered by quantum fluctuations remains of sustained interest, with proposed phases ranging from ''resonating valence bond" states \cite{anderson1, anderson2, moessner_sondhi}, to quantum dimer phases \cite{jackeli12, furukawa_gregoire, ralko_mila}, to a variety of spin liquid phases \cite{li_chen2, manuel_ceccatto, jolicoeur_bacci, hu_sheng, zhu_white, iqbal_becca, misguich_waldtmann, mishmash_xu, wietek_lauchli}, some of which may realize long-range entanglement. However, finding pristine material systems to match many of these models remains an outstanding challenge, as effects in real materials such as orbital ordering \cite{pen_sawatzky, mcqueen_cava, khomskii_mostovoy}, Jahn-Teller distortions \cite{mostovoy_khomskii, giot_alexandros}, anisotropic exchange \cite{coldea_tylczynski, vanwell, ono_mei, alicea_matthew}, and exchange disorder \cite{zhang_mourigal, sheng_henley, kimchi_senthil, zhu_chernyshev} can either induce order, quench the moments entirely, or reduce the dimensionality to quasi-1D.  

Studies of several classes of rare earth oxides have recently shown that model planes of equilateral triangles of $4f$ moments form in a high symmetry setting ideal for studying quantum fluctuations within this system. Rare earth lanthanides with the YbFe$_2$O$_4$ structure-type (YbMgGaO$_4$ \cite{li_chen, li_chen2, li_zhang, li_zhang2, li_zhang3, li_zhang4, paddison_mourigal, shen_zhao, xu_li} and TmMgGaO$_4$ \cite{li_gegenwart, cevallos_cava, shen_zhao_TMGO}) realize a spin Hamiltonian with spin-orbit entangled Ln$^{3+}$ moments (Ln=lanthanide) and anisotropic exchange couplings. At low temperatures, quantum spin liquid states have been proposed \cite{li_chen2, maksimov_chernyshev}, and experiments suggest the absence of long-range magnetic order in select compounds. The archetypal material in this class is YbMgGaO$_4$, originally proposed as a quantum spin liquid (QSL) candidate due its lack of long-range magnetic order and the observation of a low-temperature continuum of magnetic excitations; however the influence of innate chemical disorder within its lattice complicates this interpretation. 

Disorder in YbMgGaO$_4$ arises from two equally intermixed Mg$^{2+}$:Ga$^{3+}$ layers that reside between the trivalent Yb-ion triangular sheets. This alters the local chemical environment about the YbO$_6$ octahedra and introduces exchange disorder, and some recent studies have suggested the formation of a weakly-bound spin glass ground state that freezes at low temperatures \cite{zhu_chernyshev, ma_wen}. However, competing interpretations of YbMgGaO$_4$ have also proposed that the chemical disorder may enhance quantum fluctuations in favor of a spin liquid state\cite{AdvQTech2}. Other experiments failed to resolve a true spin freezing in this system \cite{PRL122, NatComm8} and are instead consistent with persistent fluctuating valence bonds at the lowest temperatures measured. While the exact nature of its magnetism remains in dispute, studies of YbMgGaO$_4$ and related variants in the YbFe$_2$O$_4$ materials family have provided a clear rational behind searching for quantum disorder in strongly spin-orbit coupled \jeff Ln-ion systems on the triangular lattice.

Building from this, an alternate family of the form $ALnX_2$ ($A =$ alkali ion, $Ln =$ Lanthanide ion, $X =$ chalcogen anion) has been identified.  These compounds are comprised of triangular planes populated with trivalent $Ln$ ions, and experiments have reported that select variants host a quantum disordered magnetic ground state \cite{liu_zhangAMX2, paper1NYO, ding_tsirlin, ranjith_baenitz, xing_sefat, xing_sefat2, xing_sefat3, ranjith_baenitz2, sarkar_klauss, sichelschmidt_doert, baenitz_doert}. The key appeal of members of this family forming in the $\alpha$-NaFeO$_2$ structure-type is their realization of an ideal triangular lattice of $Ln$ ions absent the interstitial chemical disorder. 

Specifically, \nyo \cite{paper1NYO, ding_tsirlin, ranjith_baenitz} (Figure \ref{fig:F1}) is an intriguing candidate to realize a quantum disordered ground state. NaYbO$_2$, like YbMgGaO$_4$, crystallizes in the $R\bar{3}m$ space group and contains equilateral triangular layers comprised of trigonally-distorted ($D_{3d}$ local point group) YbO$_6$ octahedra. However, in NaYbO$_2$, the in-plane Yb-Yb distance is $\sim$ 3.35 \AA \ at 1.6 K versus $\sim$ 3.4 \AA \ in YbMgGaO$_4$ \cite{li_zhang, li_zhang2, li_zhang3, li_zhang4, paddison_mourigal, shen_zhao, xu_li}, resulting in an increased antiferromagnetic exchange field of  $\theta_{CW} = -10.3$ K \cite{paper1NYO} relative to $\theta_{CW} = -4$ K in YbMgGaO$_4$ \cite{li_zhang, li_zhang2, li_zhang3, li_zhang4, paddison_mourigal, shen_zhao, xu_li}.

In NaYbO$_2$, a single separating layer of Na ions renders the interlayer Yb-Yb distance close enough to be nearly equivalent with the next-nearest neighbor in-plane distance, making interplanar interactions relevant in low-temperature exchange models \cite{paper1NYO}. This differentiates NaYbO$_2$ from YbMgGaO$_4$ that contains two nonmagnetic cationic layers separating Yb ion sheets at a larger distance. The relevant physical models for NaYbO$_2$ and YbMgGaO$_4$ may differ due to this distinction. In NaYbO$_2$, the Yb sheets stack in an $ABC$ sequence such that Yb ions within a given plane will project into the center of Yb-triangles within neighboring layers such that each interplane Yb-ion is equidistant from one another \cite{paper1NYO}. Three equivalent, antiferromagnetically-coupled interplane bonds contribute to three-dimensional geometrical frustration, bringing the material out of the typically-studied regime of purely two-dimensional triangular lattice antiferromagnets. 

Recent work has shown that \nyo does not magnetically order above 50 mK in zero field and instead manifests a low-energy continuum of spin fluctuations \cite{paper1NYO, ding_tsirlin, ranjith_baenitz}. These results are suggestive of a native, quantum disordered ground state held by the $J_{eff}=1/2$ triangular lattice of Yb$^{3+}$ moments. The ability to push this magnetic lattice into a fluctuation-driven \textit{up-up-down} antiferromagnetic state under modest magnetic fields ($H\approx$ 3 T $-$ 5 T) makes it a unique platform for exploring the critical phase boundary between fluctuation-driven order and disorder. 

In this paper, we explore the evolution of low-energy magnetic excitations across this phase boundary as well as the onset of long-range magnetic order under applied magnetic field. The first part of this manuscript reviews the single-ion spin excitations associated with the intramultiplet excitations within the Yb$^{3+}$ $J=7/2$ manifold split by the $D_{3d}$ crystal field.  The ground state doublet determined is largely in agreement with recent studies \cite{ding_tsirlin}; however, our data further reveal potential exchange splitting within the crystal electric field excitation spectrum.  At lower energies, we examine the detailed field dependence of low-energy continuum of scattering about the $\bm{Q} =(1/3, 1/3, 0)$ two-dimensional antiferromagnetic ordering zone center.  

Our spin-wave calculations qualitatively capture the field evolution of dynamics endemic to the \uud phase, and an anomalous band of excitations above the single-magnon cutoff is identified in the ordered state. Upon crossing the high-field phase boundary of the ordered state, we observe the hysteretic onset of long-range order, suggesting the formation of a noncollinear ordered state and the presence of a first order, high-field phase boundary prior to entering the quantum paramagnetic regime.  Our results demonstrate a complex evolution of order under applied field in NaYbO$_2$ and unconventional spin dynamics both within the low-field quantum disordered regime as well as within the fluctuation-driven \uud state.

\section{II. Methods}

\subsection{Sample preparation} Polycrystalline samples of \nyo were produced via a solid-state reaction of Yb$_2$O$_3$ (99.99\%, Alfa Aesar) and Na$_2$CO$_3$ (99.997\%, Alfa Aesar) in a 1:1.25 molar ratio by firing at 1000\,$^{\circ}$C for three days followed by regrinding and refiring to 1000\,$^{\circ}$C for 24 hrs in air. Volatility of Na$_2$CO$_3$ during the reaction can be controlled via crucible size and reactant mass. This results in \nyo with a small excess of Na$_2$CO$_3$ (1 – 5\%) that inhibits the reformation of magnetic Yb$_2$O$_3$ into samples. Samples were stored in a dry, inert environment, and all measurements were taken with minimal sample exposure to the atmosphere. 
  
\subsection{Neutron scattering} 

Low-energy inelastic neutron scattering data were collected on 8g of NaYbO$_2$ powder the Disc Chopper Spectrometer (DCS) instrument at the NIST Center for Neutron Research, National Institute of Standards and Technology (NIST) and the Cold Neutron Chopper Spectrometer (CNCS) instrument at the Spallation Neutron Source, Oak Ridge National Laboratory (ORNL). A 10 T magnet, dilution insert, and incident neutrons with an incident energy $E_i=3.27$ meV in the medium-resolution chopper setting were used at DCS. A 7 T magnet, dilution insert, and incident neutrons of $E_i$ = 3.32 meV were used for experiments at CNCS. Magnet background at CNCS was removed by subtracting scans collected measuring an empty copper can at 1.8 K.  High-energy inelastic neutron scattering data were obtained at the wide Angular-Range Chopper Spectrometer (ARCS) at the Spallation Neutron Source, ORNL. Data were collected on 5g of NaYbO$_2$ powder with two incident neutron energies of $E_i$ = 60 meV (Fermi 2, Fermi frequency 420 Hz)  and 150 meV (Fermi 2, Fermi frequency 600 Hz) at both 300 K and 5 K in a top loading cryostat. Background scattering from the aluminum sample can was removed from this data by obtaining data at both energies and temperatures from an empty canister. 

Elastic line analyses were conducted on DCS data by integrating $|Q|$-cuts between $E = [-0.1, 0.1]$ meV. In this data, no extra peaks due to Na$_2$CO$_3$ and superior $|Q|$-resolution allowed for the determination of magnetic Bragg reflections corresponding to the $\bm{q_1} = (1/3, 1/3, 0)$ and $\bm{q_2} = (0, 0, 0)$ ordering wave vectors previously reported \cite{paper1NYO}. The $\bm{Q} = (1/3, 1/3, 0)$ and $\bm{Q} = (1/3, 1/3, 2)$ peaks were tracked as a function of field from 0 – 10 T and 10 – 0 T by fitting data between $|Q| = [1.2, 1.29]$ \AA$^{-1}$ \ and $|Q| = [1.38, 1.52]$ \AA$^{-1}$, respectively, to Gaussian functions and extracting their integrated intensities.

\begin{figure*}[t]
	\includegraphics[scale=0.375]{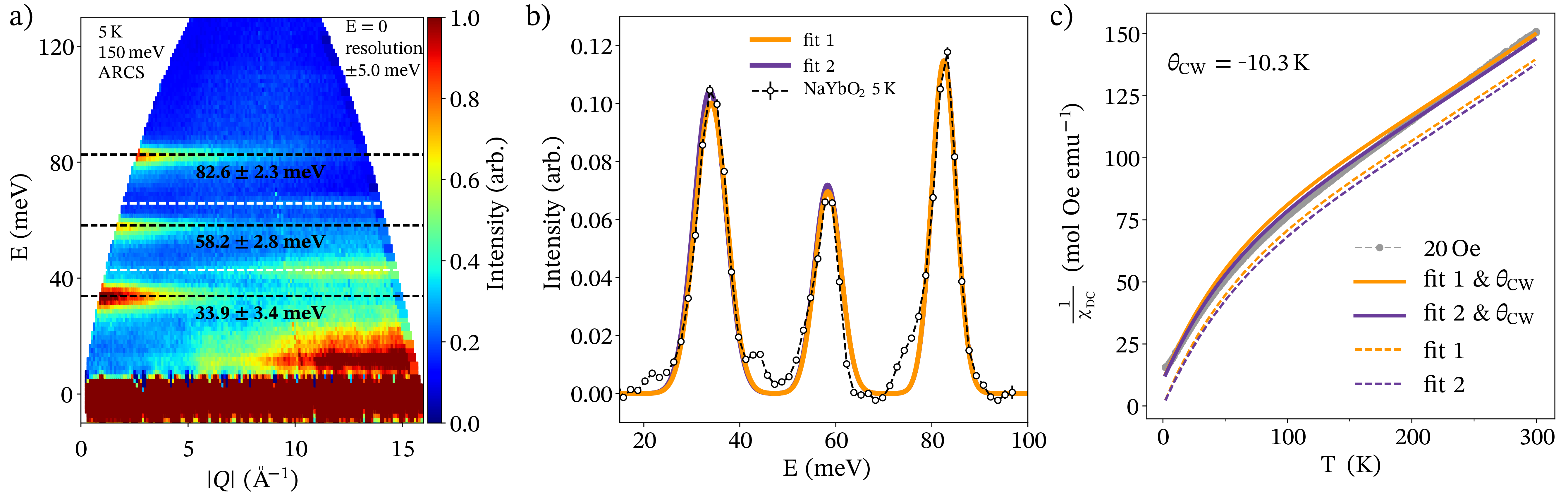}
	\caption{(a) Inelastic neutron scattering (INS) spectrum $S(Q,\hbar \omega)$ at 5 K obtained on ARCS at the Spallation Neutron Source with $E_i$ = 150 meV. The three dashed black lines feature the three CEF excitations while the dashed white lines correspond to phonons seen in integrated cuts in panel (b). Uncertainties specified next to energy values are the calculated instrument energy resolution at those energies. (b) Experimental fits to $|Q| = [3.0, 3.5]$ \AA$^{-1}$  integrated cut at 5 K of the INS spectrum. Fit 1 has the lowest $\chi^2$ value to INS data (Table 1), while fit 2 closely resembles a point charge model incorporating two coordination shells (\ref{fig:F1}). (c) Inverse magnetic susceptibility of \nyo at 20 Oe overlaid with calculated inverse magnetic susceptibilities from two comparable CEF models. The mean field exchange interaction is introduced (solid lines) to account for deviations due to strong antiferromagnetic exchange in \nyo \cite{paper1NYO}. Calculated susceptibilities that do not include the antiferromagnetic interaction (dashed lines) do not accurately reproduce the data. }
	\label{fig:F2}
\end{figure*}

\subsection{Crystalline electric field analysis} Following Hund's rules, the $4f^{13}$ Yb$^{3+}$ ions in \nyo have total angular momentum $J = 7/2$ ($L = 3$, $S = 1/2$). This 8-fold degenerate ($2J + 1 = 8$) Yb ion state is split by the $D_{3d}$ crystalline electric field (CEF) environment within the material, which, via Kramers' theorem, may be maximally split into a series of four doublets. The CEF interaction is dominated by the shell of charges closest to the central Yb ion with diminishing perturbations by shells of charges further away. Therefore, the primary splitting of the Yb$^{3+}$ manifold is due to the O$^{2-}$ ionic shell in the trigonally-compressed YbO$_6$ octahedra followed by the neighboring Yb$^{3+}$ and Na$^{+}$ cationic shells as shown in Figure \ref{fig:F1}b.

To minimize the number of terms in the CEF Hamiltonian, the $\hat{z}$ direction was chosen to align with the highest-symmetry 3-fold axis. In NaYbO$_2$, this coincides with the $c$-axis, and the CEF Hamiltonian is written with CEF parameters $B_n^m$ and Steven's operators $\hat{O}_n^m$ \cite{StevensOperators} as:

\begin{multline}
	\label{eq:CEF}
	H_{CEF} = B_2^0 \hat{O}_2^0 + B_4^0 \hat{O}_4^0 + B_4^3 \hat{O}_4^3 + B_6^0 \hat{O}_6^0 \\ + B_6^3 \hat{O}_6^3 + B_6^6 \hat{O}_6^6
\end{multline}

This Hamiltonian generates the Coulomb potential created by charges surrounding the central Yb ion in \nyo and can include multiple ionic shells. Diagonalizing the Hamiltonian returns relative CEF levels $E_0, E_1, E_2, E_3$ and eigenvectors $\phi_0^{\pm}, \phi_1^{\pm},\phi_2^{\pm}, \phi_3^{\pm}$. The $g$-tensor components of the ground state Kramers doublet are calculated with the ground state wave functions and Land\'e $g$-factor ($g_{J} = 8/7$ for Yb$^{3+}$) by:

\begin{equation}
	\label{eq:gfac1}
	g_c = 2 g_J |<\phi_0^{\pm}|J_z|\phi_0^{\pm}>|
\end{equation}

\begin{equation}
	\label{eq:gfac2}
	g_{ab} = g_J |<\phi_0^{\pm}|J_{\pm}|\phi_0^{\mp}>|
\end{equation}

When $T << E_1$, the relative intensity of the $i^{th}$ level from the ground state doublet is given by:

\begin{equation}
	\label{eq:intensity}
	\sum_{J_x,J_y,J_z, \pm, \mp} <\phi_i^{\pm,\mp}|\{J_x,J_y,J_z\}|\phi_0^{\pm}>^2  
\end{equation}

The crystal field parameters $B_n^m$ for \nyo in equation \ref{eq:CEF} can first be estimated within a simple point charge model and then established with fits to energy cuts through inelastic neutron scattering data. While point charge models are rarely accurate, they often provide as a physically-grounded starting point to determine $B_n^m$ parameters when fitting limited scattering data. The point charge model for \nyo was calculated with the following formula as implemented in the crystal field interface of Mantid Plot \cite{Mantid}:

\begin{equation}
	\label{eq:PC}
	B_n^m = \frac{4 \pi}{2n+1} \frac{|e|^2}{4 \pi \epsilon_0} \sum_i \frac{q_i}{r_i^{n+1}}a_0^n <r^n> Z_n^m(\theta_i,\phi_i)
\end{equation}

The point charge model takes into account the polar location $(r_i,\theta_i,\phi_i)$ of the $i^{th}$ charge $q_i$ relative to the central Yb ion where $<r^n>$ is the $n^{th}$ order expectation value of the Yb radial wave function, $Z_n^m$ is a spherical tesseral harmonic, $\epsilon_0$ is the permitivity of free space, $|e|$ is the elemental charge, and $a_0$ is the Bohr radius. Point charge calculations including one coordination shell ($\le$ 3.0 \AA; O$^{2-}$ ions) and two coordination shells ($\le$ 3.5 \AA; O$^{2-}$, Yb$^{3+}$, Na$^{+}$ ions) are shown in Table \ref{tab:T2}.

Models were evaluated utilizing a combination of the crystal field interface of Mantid Plot \cite{Mantid}, SPECTRE \cite{spectre}, and numerical error minimization that combined scattering data with magnetic measurements. A general minimization procedure \cite{li_zhang4, gaudet_gaulin} follows this loop:

\begin{enumerate}
	\item Initialize the Hamiltonian with a guess of CEF parameters, either from related compounds (e.g. YbMgGaO$_4$ \cite{li_zhang4}) or a point charge model.
	\item Diagonalize the CEF Hamiltonian \ref{eq:CEF}, obtain energy eigenvalues, ground state wave functions, $g$-factors, and relative intensity ratios.
	\item Calculate: $\chi^2_{tot} = \chi^2_{energies} + \chi^2_{g} + \chi^2_{intensities}$,  \\
	where: $\chi^2 = \sum \frac{(obs - calc)^2}{calc}$ and $\chi^2_{g}$ reflects the deviations of calculated $g$-factors, $\chi^2_{intensities}$ reflects the deviation of calculated intensities of transitions, and $\chi^2_{energies}$ reflects deviations of calculated energy levels.

	\item Modify CEF parameter(s) and reiterate. Accept the new parameters if $\chi^2_{tot, new} < \chi^2_{tot, old}$.
\end{enumerate}

This sequence was then iterated to obtain a global minimum that best represents the observed data.

\subsection{Spin wave analysis}

We calculate the dynamic spin structure factor $\mathcal{S}(\bm{q},\omega)$ as a function of energy and momentum. The detailed derivation has been given in \cite{paper1NYO}, and here we only outline the procedures. Starting from the spin model \eqref{Hamiltonian_J_D_B} ($D=0$), we numerically minimize the energy to obtain the classical ground state, and then use the standard Holstein-Primakoff method to construct a spin wave Hamiltonian, the diagonalization of which gives the magnon modes. A spin-spin correlation function is then readily obtained, whose momentum Fourier transform gives the desired quantity $\mathcal{S}(\bm{q},\omega)$. Note in order to simulate the powder sample results, we take average over all momentum directions and then all magnetic field orientations, and we denote the final result after the two-step average as $\overline{\overline{\mathcal{S}}}(Q,\omega)$. Note the two averages are done completely \emph{independently}. Admittedly, such a simulated spin structure factor does not fully mimic the experimental situation, since in reality the measurement just amounts to a one-time average over the grain orientations of the powder sample, and during the averaging process the rotations of the momentum and magnetic field are \emph{locked}. However, our choice of averaging is justified by the robust spectral features observed in a large region with easy-plane near-Heisenberg exchange. 

\section{III. Experimental Results}

\subsection{High-energy crystalline electric field excitations}

\begin{figure}[t] 
	\includegraphics[scale=0.45]{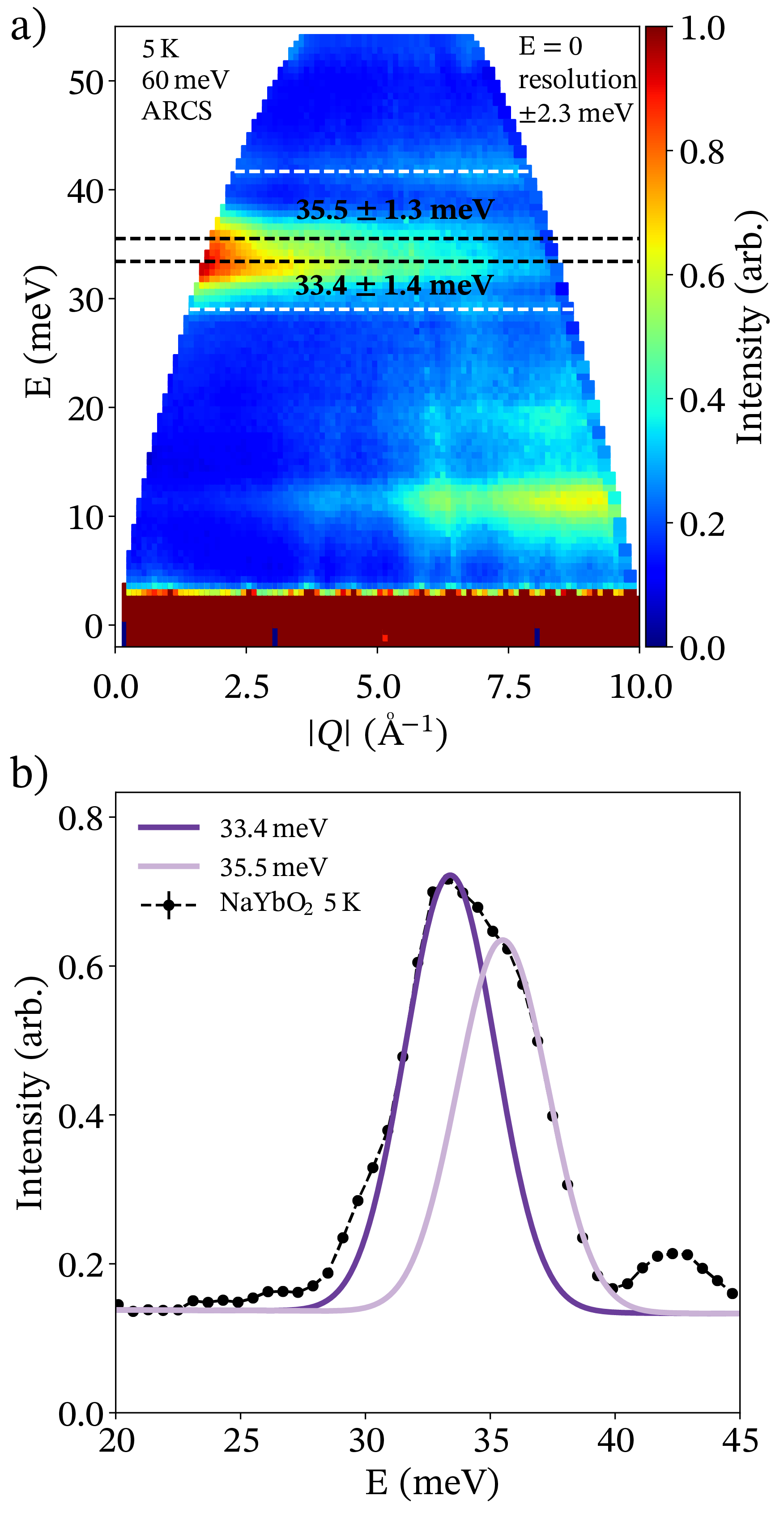}
	\caption{(a) Inelastic neutron scattering (INS) spectrum $S(Q,\hbar \omega)$ at 5 K obtained on ARCS at the Spallation Neutron Source with $E_i$ = 60 meV. 
		The two dashed black lines indicate the split peak observable at $E_i$ = 60 meV, and the dashed white lines correspond to nearby phonons. Uncertainties specified next to energy values are the calculated instrument energy resolution at those energies.  (b) $|Q|$ integrated cut of the 5 K INS spectrum from $[2.5, 3]$ \AA$^{-1}$  illustrates the first CEF excitation in \nyo is broadened by exchange-induced CEF dispersions. Fits to two resolution-limited Gaussian peaks separated by 2.1 meV accurately capture the splitting. Instrumental resolution denoted by error bars beneath each peak.} 
	\label{fig:F3}
\end{figure}

\begin{table*}[]

	\caption{Point charge (PC) models and CEF fits for \nyo compared to observed eigenergies and intensities from $E_i$ = 150 meV INS data and anisotropic $g$-factors \cite{paper1NYO} with corresponding CEF wavefunctions.  PC (3.0 \AA) incorporates only O$^{2-}$ ions in local YbO$_6$ $D_{3d}$ distorted octahedra into calculations (one coordinate shell), while PC (3.5 \AA) additionally includes nearest-neighbor Yb$^{3+}$ ions and Na$^{+}$  ions (two coordinate shells) as shown in Figure \ref{fig:F1}. Fit 1 reproduces observables with a lower $\chi^2$ value in comparison to Fit 2. However, Fit 2 resembles the signs of PC (3.5 \AA) except $B_6^0$ which is relatively small.}
	\begin{tabular*}{17cm}{l|lllllll|l|llllll}
		\hline
		& $E_1$   & $E_2$    & $E_3$    & $\frac{I_2}{I_1}$  & $\frac{I_3}{I_1}$  & $g_c$   & $g_{ab}$  & $\chi^2$ & $B_2^0$      &  $B_4^0$        & $B_6^0$         &  $B_4^3$      &  $B_6^3$       &  $B_6^6$      \\ \hline
		PC (3.0 \AA)  & 70.5 & 146.4 & 191.5 & 0.037 & 0.033 & 7.86 & 0.36 & 342.0  &-5.0675 & 0.016956 & 0.00015465 & -0.64149 & -0.00034913 & 0.0014353      \\
		PC (3.5 \AA)   & 19.2 & 35.8 & 87.5  & 0.588 & 0.061 & 0.83 & 3.44 & 16.5 & 1.6302 & 0.020578 & 0.00017436 & -0.66203 & -0.00022040 & 0.0017501       \\
		Fit 1  & 34.0 & 58.4  & 82.5  & 0.579 & 0.880 & 1.70 & 3.38 & 0.005 & -0.79877 & 0.00085658 & 0.0028000 & 0.51143 & 0.011036 & 0.015580     \\
		Fit 2 & 33.8 & 58.3  & 82.5  & 0.574 & 0.845 & 1.68 & 3.48 & 0.012  & 0.28257 & 0.0058508 & -0.00055392 & -0.76448 & -0.010493 & 0.025079  \\ \hline 
		Observed   & 33.9 & 58.2 & 82.6 & 0.548 & 0.866 & 1.72 & 3.30 & & &  &  &  &  &  
	\end{tabular*}

	\hskip+0.1cm
	\begin{tabular*}{17.025cm}{ll}
		\hline
		Fit 1:     &  \\
		$| \omega_{0, \pm}\rangle =$ & $-0.029|\mp1/2\rangle \mp0.496|\pm1/2\rangle -0.578|\mp5/2\rangle \pm0.034|\pm5/2\rangle \mp0.038|\mp7/2\rangle +0.646|\pm7/2\rangle$ \\
		$| \omega_{1, \pm}\rangle =$ & $-0.001|\mp1/2\rangle \mp0.473|\pm1/2\rangle +0.805|\mp5/2\rangle \mp0.002|\pm5/2\rangle \mp0.001|\mp7/2\rangle +0.358|\pm7/2\rangle$  \\
		$| \omega_{2, \pm}\rangle =$ & $\pm0.025|\mp1/2\rangle +0.727|\pm1/2\rangle \pm0.128|\mp5/2\rangle -0.004|\pm5/2\rangle -0.023|\mp7/2\rangle \pm0.673|\pm7/2\rangle$ \\
		$| \omega_{3, \pm}\rangle =$ & $ \pm0.614|\mp3/2\rangle +0.789|\pm3/2\rangle$ \\ \hline
		Fit 2:     &  \\
		$| \omega_{0, \pm}\rangle =$ &  $0.075|\mp1/2\rangle \pm0.582|\pm1/2\rangle -0.528|\mp5/2\rangle \pm0.068|\pm5/2\rangle \mp0.078|\mp7/2\rangle +0.604|\pm7/2\rangle$\\
		$| \omega_{1, \pm}\rangle =$ &  $1|\pm3/2\rangle$\\
		$| \omega_{2, \pm}\rangle =$ & $\mp0.574|\mp1/2\rangle -0.034|\pm1/2\rangle \mp0.047|\mp5/2\rangle +0.803|\pm5/2\rangle +0.149|\mp7/2\rangle \mp0.009|\pm7/2\rangle$ \\
		$| \omega_{3, \pm}\rangle =$ & $-0.004|\mp1/2\rangle \mp0.570|\pm1/2\rangle +0.263|\mp5/2\rangle \mp0.002|\pm5/2\rangle \mp0.005|\mp7/2\rangle +0.779|\pm7/2\rangle$
	\end{tabular*}
	\label{tab:T2}

\end{table*}

Inelastic neutron scattering (INS) data collected at $T=5$ K with an $E_i$ = 150 meV are shown in Figure \ref{fig:F2}a, revealing three CEF excitations out of the ground state doublet. The lowest-lying excitation is centered at $E_1=33.9$ meV, which is consistent with a well-separated ground state Kramers doublet, and the second and third excited states are observed centered at $E_2=58.2$ and $E_3=82.6$ meV, respectively. CEF excitation energies, integrated intensity ratios, and $g$-tensor components \cite{paper1NYO} are displayed with point charge models and CEF fits to integrated $S(\bf{Q},\hbar \omega)$ cuts of the data in Table \ref{tab:T2}. 

Models converged toward two minima labeled as ``Fit 1" and ``Fit 2" shown in Table \ref{tab:T2}. Fit 1 is unconstrained and has the lowest global error, however we note that it deviates strongly from point charge models. If the signs of the crystal field parameters (excluding $B_6^0$) are enforced to agree with a two-shell point charge model that includes O$^{2-}$, Yb$^{3+}$, and Na$^{+}$ ions (PC 2 in Table \ref{tab:T2}), then Fit 2 provides the best solution. It should be noted that a point charge model incorporating only one coordination shell of O$^{2-}$ ions (PC 1 in Table \ref{tab:T2}) could not be optimized to represent any of the observed data. Therefore, the Coulomb environment surrounding Yb ions in \nyo is heavily influenced by ions beyond the first coordination shell, consistent with recent theoretical analysis \cite{zangeneh_hozoi}.  

Figure \ref{fig:F2} (b) shows a comparison of the magnetic susceptibility data \cite{paper1NYO} with the two models obtained from INS analysis calculated with the crystal field interface in Mantid Plot \cite{Mantid}. The susceptbility calculated from the INS fits do not fully capture the data, and instead, a modified effective susceptibility $\chi_{eff}$ that includes antiferromagnetic exchange must be introduced: $\chi_{eff} = \frac{\chi_{calc}(T)}{1 - \theta_{CW} \chi_{calc}(T)}$.  Here, the previously determined antiferromagnetic exchange interaction $\theta_{CW} = -10.3$ K \cite{paper1NYO} was used in Figure 2b. 

Additionally, the moment size of each model for the ground state doublet was determined within Mantid Plot \cite{Mantid} via adding a Zeeman interaction to the CEF Hamiltonian  to compare with the ordered moment in the \uud phase previously determined for \nyo \cite{paper1NYO}. At $B = 5$ T, the expected powder-averaged moment of Fit 1 is 1.43 $\mu_B$ and of Fit 2 is 1.49 $\mu_B$. These values are 5.15\% and 9.56\% larger than the observed moment in our previous study of 1.36(1) $\mu_B$ \cite{paper1NYO}, potentially reflective of a fluctuation-reduced moment.

With a lower $E_i$ = 60 meV, the INS spectrum at $T=5$ K only contains a single excitation out of the lowest-lying Kramers doublet as shown in Figure \ref{fig:F3}. In this higher energy resolution setup, it becomes clear that the first excited doublet consists of two asymetrically-split, resolution-limited peaks with centers at 35.5 $\pm$ 1.3 and 33.4 $\pm$ 1.4 meV. This splitting of approximately $\Delta E=2.1$ meV is too narrow to observe in the higher $E_i$ = 150 meV data of Figure \ref{fig:F2}. For comparison, instrumental resolutions at both energy transfers are tabulated in Figure \ref{fig:F1}c as errors in observed CEF excitation peak centers. Splitting of the lowest-lying doublet is naively not allowed in the CEF Hamiltonian from equation \ref{eq:CEF} by Kramers theorem and the $D_{3d}$ Yb ion point group symmetry. While such a splitting can indicate two different Yb local environments in the lattice, earlier diffraction studies do not resolve this or other modes of average chemical disorder\cite{paper1NYO}. Instead, the CEF Hamiltonian neglects any interactions beyond the single-ion level, such as exchange-induced dispersion \cite{sumarlin1994dispersion}, which is the likely origin of the splitting and a further reflection of the sizable Yb-exchange interactions in this material.  We note here though that more complicated forms of disorder in the local structure generating this CEF splitting cannot be completely excluded by the present data.

\begin{figure*}[t]
	\includegraphics[width=\textwidth*10/10]{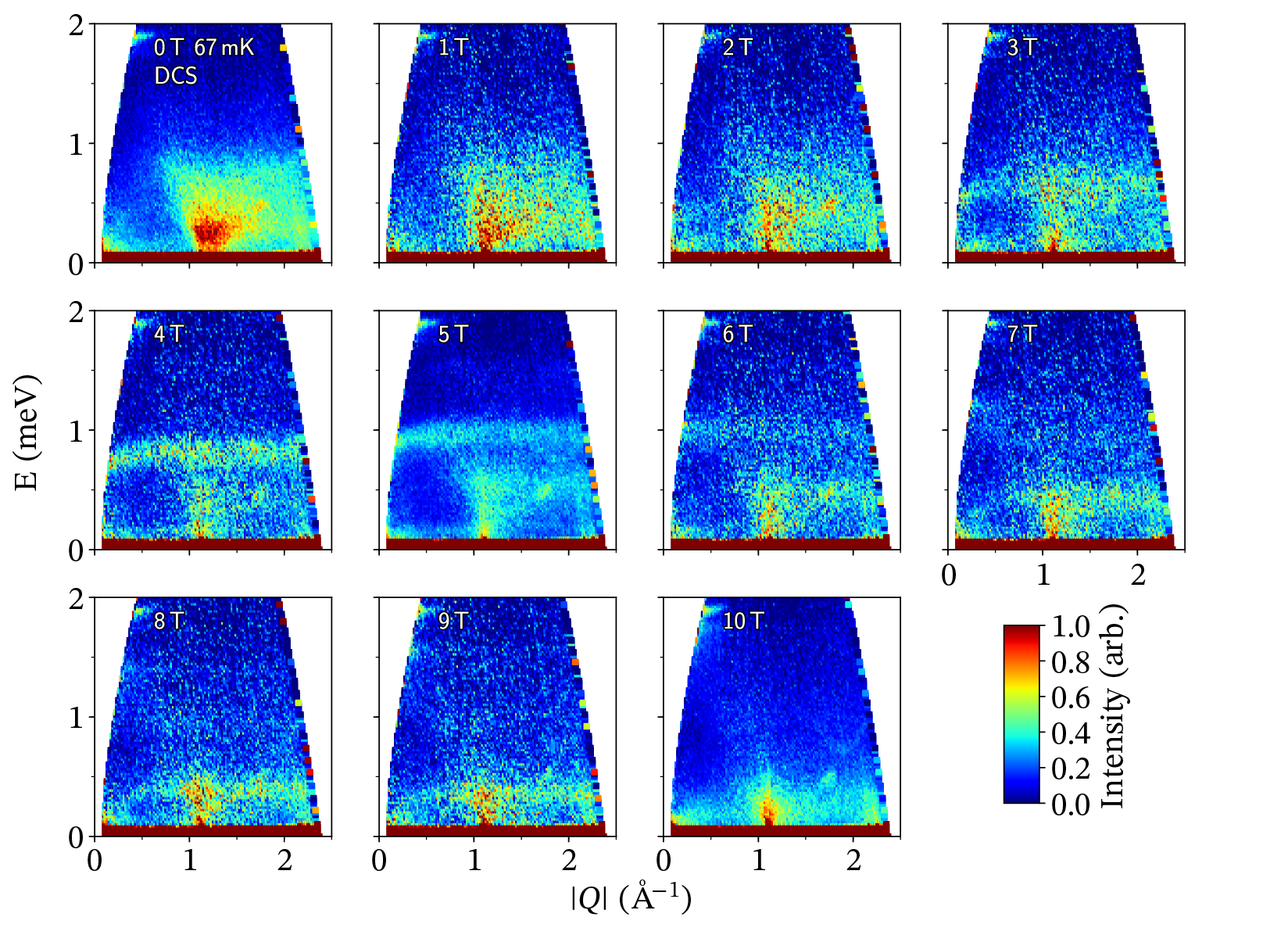}
	\caption{Low energy inelastic neutron scattering (INS) spectrum $S(Q,\hbar \omega)$ of \nyo powder at varying fields collected on DCS. With increasing field, \nyo evolves from a gapless quantum disordered ground state (0 – 2T) into an up-up-down equal moment magnetic structure (3 – 8T) and a field-polarized state at high field (9 – 10T). Data were collected with longer scans at 0, 5, and 10 T to increase resolution. Detector spurions occur at $[$0.5 \AA$^{-1}$, 1.8 meV$]$ and $[$1.75 \AA$^{-1}$, 0.4 meV$]$. Data were collected between 67 – 100 mK.}
	\label{fig:DCS}
\end{figure*}


\subsection{Low-energy magnetic excitations }

In order to investigate the correlated spin dynamics of the Yb-moments, low-energy inelastic neutron scattering measurements were performed below 100 mK. Figures \ref{fig:DCS} and supplemental Figures S1 and S2 show the field-dependent evolution of the INS spectrum of \nyo powder across a series of fields spanning from 0 T to 10 T. $H=0$, 5, and 10 T data sets were counted $\approx 8$ times longer relative to other fields in Fig. 4, and the temperature stabilized between $T=67$ to 100 mK across this field range. 

At zero-field, \nyo contains a continuum of excitations from the quantum disordered ground state that evolve into the \uud ordered phase as previously reported \cite{paper1NYO}. The diffuse continuum is centered about the two-dimensional magnetic zone center $\bm{Q} =(1/3, 1/3, 0)$ with a bandwidth of approximately 1 meV. With increasing field at base temperature, the spectral weight condenses and splits, with part of it coalescing into the elastic line and part of it pushed upward within a nearly flat, powder-averaged band near 1 meV in the ordered state (Figure \ref{fig:DCS}). Upon exiting the ordered state at 10 T, the remaining resolvable scattering in this energy window primarily resides above the two-dimensional magnetic zone center $\bm{Q} =(1/3, 1/3, 0)$  ($|Q| =$ 1.25 \AA$^{-1}$) and the $(0,0,3)$ Bragg peak ($|Q| =$ 1.15 \AA$^{-1}$). 

Figure 5 further parameterizes the spectral shift under field via momentum-averaged, energy cuts.  In Fig. 5a the low-energy spectral weight endemic to the 0 T quantum disordered ground state can be seen to diminish and pushes upward in energy with increasing field.  Data integrated about the two-dimensional magnetic zone center $\bm{Q} =(1/3, 1/3, 0)$ show a pile-up of spectral weight near $E_{peak}=0.25$ meV in the quantum disordered phase which continuously shifts upward upon approaching the ordered phase. Prior heat capacity data \cite{paper1NYO} demonstrate that gapless excitations necessarily persist below this peak in the fluctuation spectrum, and it is reminiscent of deconfined spinons coexisting with fluctuations associated with short-range antiferromagnetic correlations simulated in variational Monte Carlo studies of triangular lattice models \cite{ferrari2019dynamical}.

Looking at higher energies, Fig. 5b illustrates the upward shift of the flat band of powder-averaged modes associated with \uud order.  Increasing field pushes the high frequency mode upward consistent with linear spin wave calculations discussed later in this paper and with prior nonlinear spin wave treatment of the \uud state \cite{kamiya2018nature}.  Upon approaching the upper-field phase boundary for the \uud state, this band broadens and diminishes, while a second lower band appears near $E=0.5$ meV and shifts downward in energy with increasing field.  This suggests a softening of at least one branch of modes within the \uud state toward the upper field boundary and disappearance of magnetic order.  It may also presage the existence of an upper field phase boundary into another ordered state. 

A third, anomalous, feature appears in energy cuts through the $\bm{Q} =(1/3, 1/3, 0)$ position in Fig. 5a.  Upon entering the \uud phase, spectral weight appears at an energy near $E=1.5$ meV which is $\approx 3J$ and far above the expected single-magnon cutoff.  This peak, which we label $E^*$, appears in the fully formed \uud state \cite{paper1NYO} and suggests substantial spectral weight within the multimagnon scattering channel or an unconventional spin wave mode pulled out of the continuum \cite{mourigal_zhitomirsky}.  If the origin is two-magnon scattering, substantial weight present in the longitudinal channel of spin fluctuations indicates strong magnon-magnon interactions in this material and would be consistent with the reduction in the ordered moment due to strong quantum fluctuations.

\begin{figure}[t]
	\includegraphics[scale=0.5]{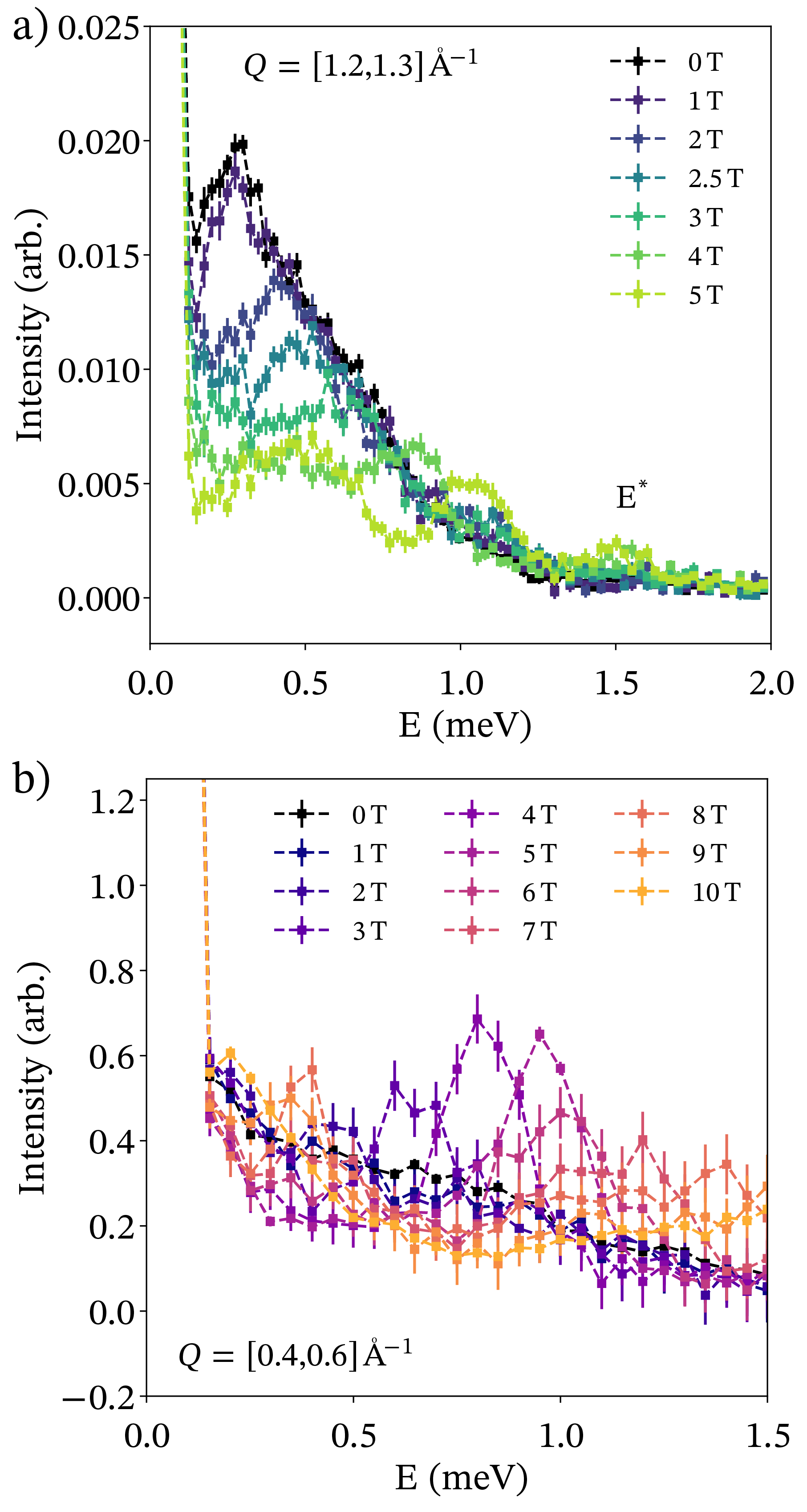}
	\caption{(a) Low energy inelastic neutron scattering (INS) of \nyo from CNCS at 5 T plotted on a logarithmic scale showing a weak third magnetic peak ($E^*$) centered at 1.5 meV above the flat, powder-averaged bands from the \uud phase. The third magnetic mode arises from a multi-magnon process convolving the two lower-energy bands of the \uud phase. At 5 T, this multi-magnon band is readily visible as it is well-separated from the powder-averaged \uud structure. (b) Low $|Q|$-integrated energy cut of INS data of \nyo from DCS data (\ref{fig:DCS}) showing the evolution of the flat \uud mode as a function of field. }
	\label{fig:F7}
\end{figure}

\subsection{Field-dependent magnetic order}

\begin{figure*}[t]
	\includegraphics[width=\textwidth*10/10]{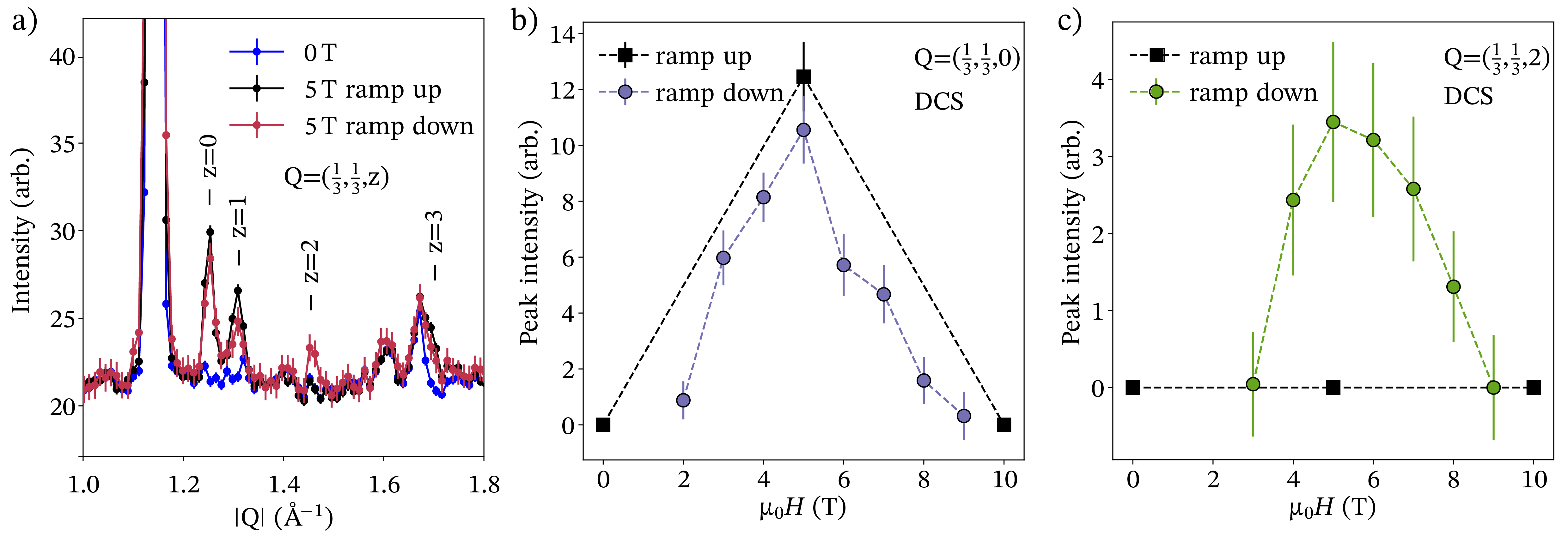}
	\caption{(a) Neutron powder diffraction data collected from an elastic line cut on the DCS instrument at the NIST Center for Neutron Research with 5 \AA incident neutrons between $E = [-1,1]$ meV. In zero field, no new magnetic reflections arise.  0T data was collected at 67 mK and 5T data was collected at 74 mK (b-c) Integrated intensity of the  $\bm{Q} = (1/3, 1/3, 0)$ and $(1/3, 1/3, 2)$  peaks respectively upon sweeping field from 0 $\rightarrow$ 10 T and 10 $\rightarrow$ 0 T between 67 mK (0T) and 92 mK (10T).  Data were analyzed via Gaussian fits to elastic line DCS data. At 0 and 10 T, there was no observable intensity at the  $\bm{Q} = (1/3, 1/3, 0)$ position, indicating \nyo is in its quantum disordered ground state and field-polarized state, respectively. No quantifiable intensity is ascribable to the  $\bm{Q} = (1/3, 1/3, 2)$  peak on field-ramping from 0 $\rightarrow$ 10 T. Though, upon ramping downward 10 $\rightarrow$ 0 T, the $\bm{Q} = (1/3, 1/3, 2)$ peak appears between 9 and 3 T.}
	\label{fig:F6}
\end{figure*}

Upon analyzing the elastic line ($E=0$ meV) in Figure \ref{fig:F6}, the onset of static magnetic order in \nyo under applied field can be observed. At 5 T, peaks appear at momentum transfers corresponding to the $\bm{Q} =(1/3, 1/3, z)$ positions as shown in Fig. 6a.  The onset and subsequent exit of three-sublattice order upon cycling the magnetic field upward from 0 $\rightarrow$ 10 T can be monitored via the $\bm{Q} =(1/3, 1/3, 0)$ reflection shown in Figure \ref{fig:F6}.  This is the strongest magnetic reflection that does not coincide with a structural Bragg peak, and it is present in both the collinear \uud state as well as in other noncollinear states such as the canted V-state and Y-state order detailed in Section IV of this paper. The intensity of this peak is shown in Figure \ref{fig:F6}b for a field-ramp up at $H=0$, 5, and 10 T and for a field-ramp down from 9 T to 2 T. Data points at 0 and 10 T are placed for reference, as no integrable intensity is present at those fields. As previously determined with A.C. susceptibility \cite{paper1NYO}, the powder-averaged boundary of field-induced magnetic order resides between 2 – 3 T and 8 – 9 T which coincides with the onset and disappearance of the  $\bm{Q} = (1/3, 1/3, 0)$ peak.

Surprisingly, after cycling the \nyo powder to 10 T and then returning down to 5 T, the magnetic order stabilized changes.  Specifically, the $\bm{Q} =(1/3, 1/3, 2)$ peak, previously absent on cycling the field upward and whose suppression is indicative of collinear \uud order, appears upon cycling the field downward. This new peak coincides with the onset of the  $\bm{Q} =(1/3, 1/3, 0)$ peak, and the field evolution of the  $\bm{Q} = (1/3, 1/3, 2)$ peak is displayed in Figure \ref{fig:F6}c. This hysteresis in the onset of magnetic order upon traversing the upper-field phase boundary from above suggests a first-order phase line and the appearance of a noncollinear ordered state.  Recent calculations have predicted such a phase boundary into a coplanar V-state in triangular lattice systems with strong interplanar exchange coupling \cite{yamamoto2015microscopic}; however identifying the precise spin structure trapped across this first-order phase line requires future single crystal measurements.  


\section{IV. Theoretical analysis}

\subsection{Classical 2D XXZ model}
In this section we further the study of the classical phase diagram of the 2D XXZ model on the triangular lattice in the presence of field along arbitrary direction. The Hamiltonian reads
\begin{equation}\label{Hamiltonian_J_D_B}
\begin{split}
H = \sum_{\langle i,j\rangle} J_z S^z_iS^z_j+J_{xy}( S^x_iS^x_j+S^y_iS^y_j)\\
 + D \sum_i (S^z_i)^2 -\sum_i \mu_B g_{\mu\nu} B^\mu S^\nu_i,
\end{split}
\end{equation}
where $\mu,\nu=x,y,z$, $g_{\mu\nu} = \text{diag}(g_{xy},g_{xy},g_z)$. We have shown in \cite{paper1NYO} that the four-term Hamiltonian representing the 2D XXZ model reduces to this one, assuming three-sublattice order. The main conclusions from our previous study \cite{paper1NYO} of this Hamiltonian are: (1) the ground state is a three-sublattice 120$^{\circ}$ structure that (2) evolves into a canted phase with an external field, which becomes a canted up-up-down structure depending on the field strength and direction, and  (3) a good fit to the inelastic neutron scattering powder-averaged spectrum of NaYbO$_2$ is produced at $J_{xy} = 0.51$ meV and $J_z = 0.45$ meV.

We now use $d=a,b,c$ to label the three sublattices, and define $S^\mu_d = S n^\mu_d$, where $\bm{n}_d$ is a unit vector. We further define $\Sigma^\mu = \sum_d n^\mu_d$, then the classical ground state is obtained by minimizing the following quantity
\begin{align}
E = \frac{H}{3NS^2 J_{xy}/2}=&A(\Sigma^z-h_z)^2+(\Sigma^x-h_x)^2  \nonumber \\
    &+(\Sigma^y-h_y)^2 -\delta \sum_d (s^z_d)^2-C,
\end{align}

where $N$ is the number of sites in the 2D lattice, and we have defined 

\begin{align}
A = \frac{J_z}{J_{xy}},\quad (h_x,h_y,h_z) = \frac{\mu_B}{3S}\left(\frac{g_{xy} B^x}{J_{xy}},\frac{g_{xy} B^y}{J_{xy}},\frac{g_z B^z}{J_z}\right), \nonumber \\
\delta = A-1-\frac{2D}{3J_{xy}},\quad
C=\frac{\frac{H^2_z J_{xy}}{3J^2_z}+\frac{H^2_{xy}}{3J_{xy}}-9 J_{xy}S^2}{3 S^2J_{xy}}.
\end{align}

We then write $\bm{n}_d = (\sin \theta_d \cos \phi_d,\sin \theta_d\sin\phi_d,\cos \theta_d)$ and take the derivatives with respect to angular variables,
\begin{subequations}
	\begin{eqnarray}
	\frac{\partial E}{\partial \theta_d}=0&\Rightarrow&
	-2A(\Sigma^z-h_z) s^{xy}_d
	+2(\Sigma^x-h_x)  s^z_d\frac{s^x_d}{s^{xy}_d}\\
	&&+2(\Sigma^y-h_y) s^z_d\frac{s^y_d}{s^{xy}_d}
	+2\delta s^z_d s^{xy}_d
	=0,\label{eq_for_theta}\\
	\frac{\partial E}{\partial \phi_d}
	=0&\Rightarrow&-2(\Sigma^x-h_x)s^y_d+2(\Sigma^y-h_y)s^x_d=0,\label{eq_for_phi}
	\end{eqnarray}
\end{subequations}
where $s^{xy}_d = (s^x_d)^2+(s^y_d)^2$ (we will use similar notation for other quantities). We then have the following two cases:\\

	\noindent \textit{Case 1:} If $\Sigma^x-h_x$ and $\Sigma^y-h_y$ do not vanish at the same time: suppose $\Sigma^y-h_y\neq 0$, then we have $\frac{\Sigma^x-h_x}{\Sigma^y-h_y}=\frac{s^x_d}{s^y_d}$, the order is coplanar in the plane containing $z$ axis. Therefore we are actually minimizing
	\begin{equation}\label{coplanar_energy}
	E_{\text{coplanar}}
	=A(\Sigma^z-h_z)^2+(\Sigma^{xy}-h_{xy})^2-\delta \sum_d (s^z_d)^2.
	\end{equation}
	This will be treated in detail below. \\ \\
	\noindent \textit{Case 2:} Otherwise, we have
	\begin{equation}\label{sxhxsyhy}
	\Sigma^x-h_x=\Sigma^y-h_y=0.
	\end{equation}
	Plugging this into Eq.~\eqref{eq_for_theta}, we see that 
	\begin{equation}
	s^{xy}_d\left[\delta s^z_d - A(\Sigma^z-h_z)\right]=0,
	\end{equation}
	has four cases, depending on how many $s^{xy}_d=0$; note that
	if two or all three $s^{xy}_d=0$ then the situation is included in the first case. Therefore we only need analyze two possibilities.

		The first is if $s^{xy}_d\neq 0$ for all $d=a,b,c$, then we must have 
		\begin{equation}\label{equalszcondition}
		s^z_a=s^z_b=s^z_c=\frac{Ah_z}{\delta-3A},
		\end{equation}
		In other words, the spins have equal $z$ component. We call this the ``canted-I'' phase. This solution should be considered only when $s^z=|\frac{Ah_z}{\delta-3A}|\leq 1$ and $3\sqrt{1-(s^z)^2}\leq h_{xy}$, i.e.
		\begin{equation}\label{eq:oval1}
		\frac{h^2_{xy}}{9}+\frac{h^2_z}{(3-\delta/A)^2} \leq 1.
		\end{equation}
		The ground state manifold is a degenerate 1D parameter space, resulting from the different ways the $xy$ in-plane vectors satisfy Eq.~\eqref{sxhxsyhy}. Note that in this case the three equations for $\theta_d$ are all independent, but the three equations for $\phi_d$ are reduced to just two equations. Therefore we should get a 1D degenerate classical ground state manifold. 
		
		The second possibility is to suppose $s^{xy}_a=0$ and $s^{xy}_b=s^{xy}_c\equiv s^{xy}\neq 0$, then we must have $s^z_b=s^z_c$. We call this the ``canted-II'' phase. In this case there are already four equations therefore the spins are uniquely determined, leaving no classical ground state degeneracy. This solution should be considered only when $s^{xy} =\sqrt{1-(s^z)^2} \geq h_{xy}/2$, i.e.
		\begin{equation}
		\frac{h^2_{xy}}{4}+\frac{(s^z_a-h_z)^2}{(2-\delta/A)^2}\leq 1,\qquad s^z_a=\pm1.
		\end{equation}

In summary, the classical ground state of the Hamiltonian \eqref{Hamiltonian_J_D_B} can only be one of the following types: coplanar (in which the order plane must contain the $z$ axis), collinear, ``canted-I'' (in which the three spins have the same $z$ component), or ``canted-II'' (in which one spin lies along $z$ and the other two have the same $z$ component). For generic field directions, only the ``canted-I'' states can form a 1D degenerate classical ground state manifold.

\subsection{Classical phase diagram}
\begin{figure}
	\centering
	\includegraphics[width=.45\textwidth]{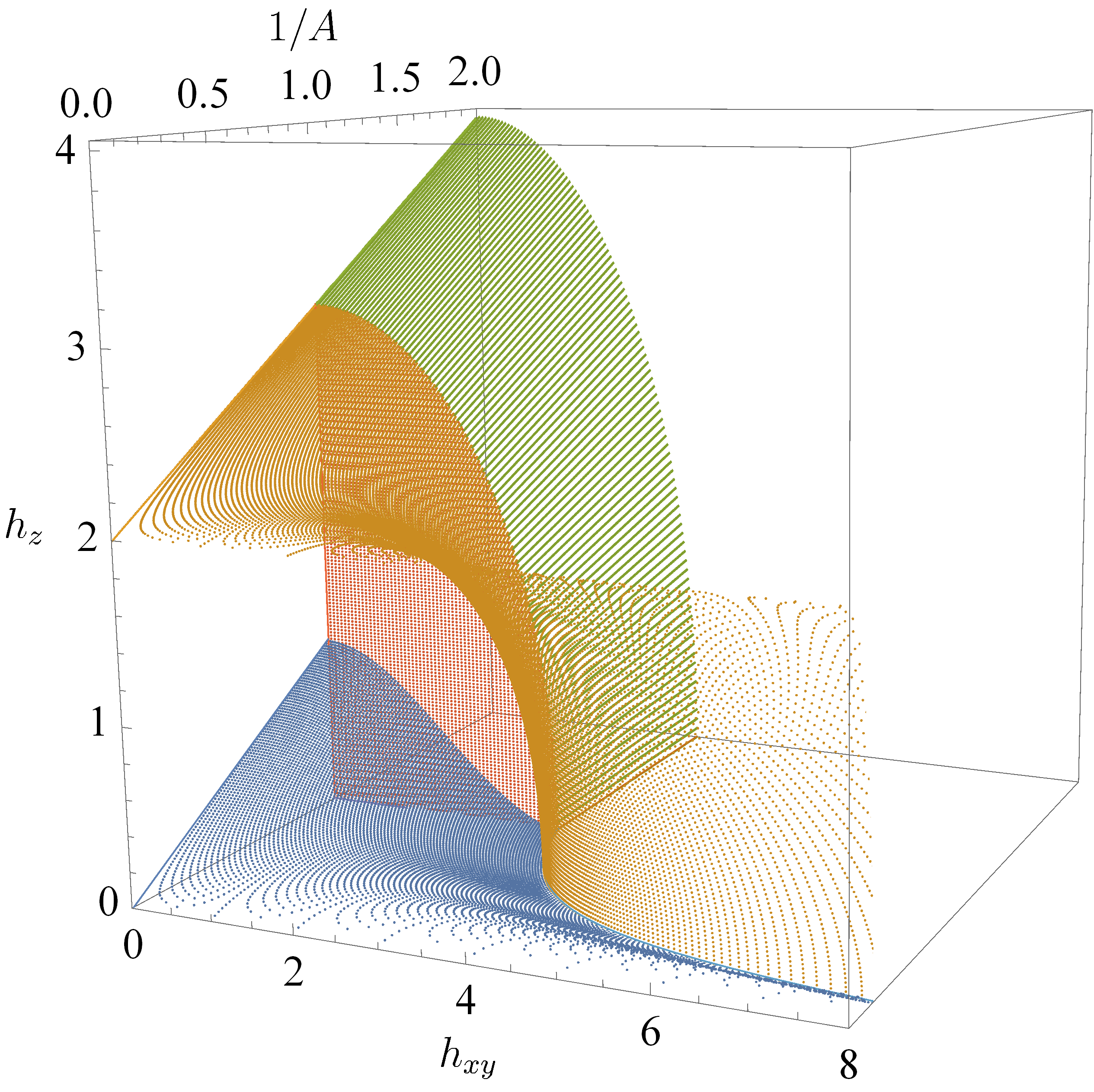}
	\caption{Classical phase diagram of the 2D XXZ model on a triangular lattice in presence of magnetic field. The 3D phase space is parametrized by $(h_{xy},A^{-1},h_z)$, where $h_{xy}=\sqrt{h_x^2+h_y^2}$. Only first octant ($h_{xy}\geq0,A^{-1}\geq0,h_z\geq0$) is considered. The blue surface separates the ``Y'' and the ``V'' phases; the red plane separates the phases between the $A<1$ and the $A>1$ regions; the green surface separates the ``canted-I'' and the ``paramagnetic'' phases in the region $A<1$, and the orange surface separates the ``V'' and the ``paramagnetic'' phases in the region $A>1$.}
	\label{fig:phases}
\end{figure}

In the following we will set the onsite ion term $D=0$, which means $\delta=A-1$. We now present a concrete phase diagram for the Hamiltonian in the $(A,h_{xy},h_z)$ phase space. For an illustration of the phase diagram, see Fig.~\ref{fig:phases}. \\ 

\noindent \textit{Easy-plane anisotropy:} In the easy-plane anisotropy region ($0<A<1$), the phase diagram has been analytically obtained \cite{paper1NYO}
	\begin{equation}
	\left\{\begin{array}{ll}
	\frac{h^2_{xy}}{9} + \frac{h^2_z}{(1/A+2)^2} \geq 1\colon & \text{``paramagnetic'' phase;}\\
	\frac{h^2_{xy}}{9} + \frac{h^2_z}{(1/A+2)^2} < 1\colon &  \text{``canted-I'' phase.}\end{array}\right.
	\end{equation}
	The phase boundary is the same as Eq.~\eqref{eq:oval1} if we take the equality. The ``paramagnetic'' phase has a unique classical ground state while in the ``canted-I'' phase the classical ground states are accidentally degenerate and form a one-dimensional manifold, subject to the constraints \eqref{sxhxsyhy} and \eqref{equalszcondition}. \\
	
\noindent \textit{Easy-axis anisotropy:} In the easy-axis anisotropy region ($A>1$), three phases exist: the ``paramagnetic'' phase, the ``Y'' phase and the ``V'' phase. We define the ``V'' phase to be such that two of the spins have identical orientation which is different from the third one, while we define the ``Y'' phase to be such that the orientation of each is different from the other two. In \cite{paper1NYO} we were able to find the boundary that separates the ``paramagnetic'' phase from the ``V'' and the ``Y'' phases, but were unable to find the phase boundary between the latter two. Here we provide the complete phase diagram:
	\begin{equation}
	\left\{\begin{array}{ll}
	\frac{h^2_{xy}}{(A+2)^2}  + \frac{h^2_z}{(1/A+2)^2} \geq 1\colon & \text{``paramagnetic'' phase;}\\ \\
	\frac{h^2_{xy}}{(A+2)^2}  + \frac{h^2_z}{(1/A+2)^2} < 1 \colon &  \text{``V'' phase;}\\
	\text{ and } h_z\geq h_{z,0}(A,h_{xy}) \\ \\
	h_z\leq h_{z,0}(A,h_{xy})\colon &  \text{``Y'' phase,}\end{array}\right.
	\end{equation}
where we have defined critical $h_{z,0}(A,h_{xy})$, which is a function of $A$ and $h_{xy}$. $h_{z,0}$ is determined from the following group of equations, taking the smallest positive nonzero solution for $h_{z,0}$:
\begin{eqnarray}\label{eq:a_c_h_z}
	A(a+c-h_z)\sqrt{1-a^2} &=& a(\sqrt{1-a^2}+\sqrt{1-c^2}-h_{xy}), \nonumber \\
	A(2a-h_z)\sqrt{1-c^2} &=& c (2\sqrt{1-a^2}-h_{xy}), \nonumber \\
	c &=& h_z - a^3(A^{-1}-1)-2a.
\end{eqnarray}
the corresponding solution for the other variables $a=n^z_a=n^z_b$ and $c=n^z_c$ gives the $z$ component of the three spins in the ``V'' phase. The first two equations simply come from the saddle point equation \eqref{eq_for_theta}; the last equation originates from the fact that, at the vicinity of the phase boundary between ``V'' and ``Y'' the energy \eqref{coplanar_energy} (note now $\delta=A-1$) takes the form $E_{\text{coplanar}}\sim \text{Const.}+\mathcal{O}(a-b)^3$, i.e. when expanding $E_{\text{coplanar}}$ in powers of $a-b$ both the first and second order terms must vanish (in fact the third order vanishes too). The analytical solution of Eqs.~\eqref{eq:a_c_h_z} to $h_z$ is hard; however, when $A$ is small enough ($A<2$ for a numerical estimation), the solution for $h_z$ can be well approximated by the empirical form
\begin{equation}
h_z = \frac{1}{A}\left(1-\frac{h_x}{2+A}\right)^2\left[b+(1-b)\left(1-\frac{h_x}{2+A}\right)\right]^2
\end{equation}
with appropriate choice of $b$ as a fitting parameter. Note in the limit $h_{xy}=0$ (perpendicular field) we recover the result $h_{z,0} = 1/A$ for the boundary between the ``Y'' and the ``V'' phases, and $h_{z,1}\equiv 1/A+2$ for the boundary between the ``V'' phase and the fully polarized phase \cite{miyashita1986magnetic}. Note also that a special type of the ``V'' state, the \uud state, should be distinguished as another distinct phase in the $h_{xy}=0$ limit, but such a phase loses its meaning as soon as an in-plane field component is turned on.

Applying these results to NaYbO$_2$, which carries easy-plane exchange couplings $J_z = 0.45\,$meV and $J_{xy}=0.51\,$meV, we are left only with two phases: the ``canted-I'' phase and the ``paramagnetic'' phase. The critical field for the onset of the ``paramagnetic'' phase is
\begin{equation}
B_{z,\text{c}} = 21.15\,\text{T},\qquad B_{xy,\text{c}} = 12.03\,\text{T},
\end{equation}
and when the field is oriented in other directions, the corresponding critical $B_c$ interpolates between these two values. 

We note here that the canted-I phase does not exactly match the experimentally reported \uud state.  A slightly canted \uud state can however form within the manifold of allowed canted-I states. This may be beyond the detection of the current powder measurements, or, alternatively, we envision that quantum fluctuations or other exchange interactions may lead to a slightly different ground state from those predicted in the purely 2D classical XXZ phase diagram.  Despite this difference, the dynamics calculated from the classical 2D model are likely to be relatively insensitive to small differences in the ordered phase such as a small degree of noncollinear canting predicted in the present model.

\subsection{Spin wave analysis}
The two-step averaged dynamic spin structure factor $\overline{\overline{\mathcal{S}}}(Q,\omega)$ for various field strengths is plotted in Fig. 8. As we expect that the ground state of \nyo is strongly renormalized by quantum fluctuations, this model only captures features in the field-induced ordered state of \nyo where quantum fluctuations in the material are suppressed. Three main features can be observed immediately:

	Region 1: Zero energy intensity at the two-dimensional magnetic zone center $\bm{Q} =(1/3, 1/3, 0)$ ($|Q| =$ 1.25 \AA$^{-1}$) can be observed for a large range of field values, indicating the existence of gapless Goldstone mode at $\Gamma$ point. The zero energy intensity is the highest at zero field with a sharp linear dispersion, and as field starts to increase, such zero energy intensity decreases, while the intensity at low but finite energy begins to develop. The zero energy intensity becomes extremely weak but still observable as the field goes beyond 13\,T, and finally vanishes entirely at high field values 22\,T. Such behavior of the gapless intensities can be understood from the classical ground state: the ground state belongs to the ``canted-I'' phase, which forms 1D degenerate ground state manifold and possesses one Goldstone mode for generic field directions and strength.  As field increases, the configuration with in-plane field first reaches critical field at 12\,T, and the structure factor of such configuration becomes gapped due to the vanishing of Goldstone modes. As field further increases, more and more configurations reach their critical field and become gapped, and at $B\sim21\,$T the last gapless configuration (corresponding to a perpendicular field) vanishes, leaving behind a fully gapped low energy intensities.
	
	Region 2: A flat intensity region is discernible at fields smaller than $\sim6\,$T. At zero field, the flat intensity appears at energy $E\sim0.8\,$meV; as field increases, the flat region starts to split and form two flat regions, one moving towards higher energy and the other towards lower energy. The higher energy flat region approaches $E\sim 1.0\,$meV at $B=5\,$T, which corresponds to the observed flat intensity in neutron scattering experiments at the same field strength. As field further increases, the higher and lower energy flat intensities vanish at $B\sim 6.5\,$T and $\sim8.5\,$T, respectively.
	
	Region 3: The behavior of the intensities at zero momentum $|Q|\sim0$ change drastically as field is varied. When the field is small, the zero momentum intensity is weak and at low energy, resulting a visual downturn from the higher energy flat intensities. As field increases, the zero momentum intensity also increases and moves towards higher energies; the downturn finally vanishes at $B=5.5\,$T, resulting a globally flat intensity across all the plotted momenta. Further increasing the field will result in an upturn of the zero momentum intensity, meaning the zero momentum intensity further increases and become the highest energy intensity in the plot. The evolution of the zero momentum intensity is closely related to the large in-plane component of the field \cite{paper1NYO}; after the field exceeds the in-plane critical field $B_{xy,\text{c}}$, the configuration with an in-plane field has a gapped spin wave spectrum, which is responsible for the highly dispersed, high intensity branch of the plot. 

\begin{figure*}[t]
	\includegraphics[width=\textwidth*10/10]{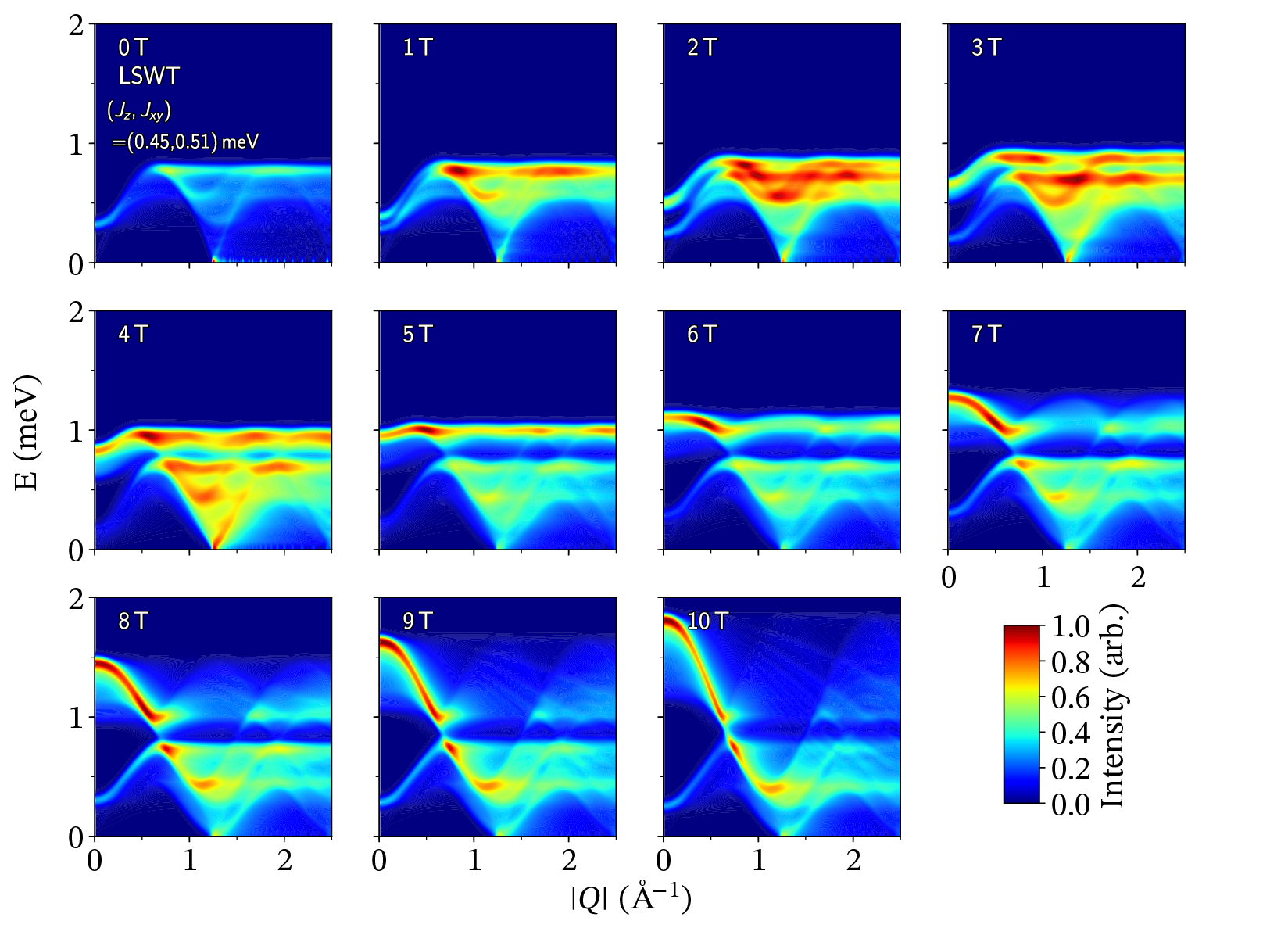}
	\caption{ Linear spin wave theory (LSWT) calculations showing $S(Q,\hbar \omega)$ as a function of field for powder-averaged Yb$^{3+}$ ions on a two-dimensional triangular lattice assuming three-sublattice ordering derived from the proposed spin model for \nyo \cite{paper1NYO}. At 0 T, \nyo does not show magnetic ordering, and therefore LSWT fails to capture the continuum of excitations from the quantum disordered ground state.  }
	\label{fig:LSWT1}
\end{figure*}

\section{V. Discussion}

Crystalline electric field calculations allow for the determination of characteristics of Yb ions in the single-ion limit. The two proposed fits for \nyo shown in Table \ref{tab:T2} and Figure \ref{fig:F2} are qualitatively comparable. Although the first fit does not have a physical point-charge basis, it better represents the observed INS data, $g$-tensor components in Table \ref{tab:T2}, as well as the observed moment size of 1.36(1) $\mu_B$ in the 5 T magnetically-ordered \uud phase \cite{paper1NYO}. Discrepencies between this model and that recently reported by Ding et al. \cite{ding_tsirlin} likely arise from the way the susceptibility was treated in modeling the data.  In the present case, susceptibility data in Fig. 2b were modeled by incorporating the large $\theta_{CW}$ field of NaYbO$_2$ \cite{paper1NYO} rather than fitting the high temperature part of the susceptibility, absent exchange.  

Within the higher incident energy INS data in Figure \ref{fig:F2}, excitations out of the CEF ground state do not show resolvable broadening effects from chemical disorder \cite{li_zhang4, gaudet_gaulin}, and all intramultiplet excitations can be indexed. Site mixing changes to CEF excitations can be roughly parametrized via the point charge model shown in equation \ref{eq:PC}, and in supplemental Table S1 varying degrees of chemical site mixing are shown incorporated out to the second shell of ions. At this distance, the smallest energy shift of a CEF excitation due to disorder in Table S1 is more than 4 meV, and the disordered CEF doublet energies do not shift equally under site disorder. Resolving these shifts and any resulting energy broadening are within instrumental resolution of the $E_i =$ 150 meV data, and the absence of broadening in the data constrains any innate chemical disorder in NaYbO$_2$ to be weak. While unresolved, local disorder can split CEF levels in certain scenarios and its influence cannot be completely excluded in our present data, prior analysis of the average structure of NaYbO$_2$ does not resolve the necessary disorder.

Instead, the subtle peak splitting observed in the first excited doublet in the higher resolution $E_i = 60$ meV of Fig. 3 likely arises from substantial exchange between Yb ions in NaYbO$_2$. Yb ions reside only $\sim$ 3.35 \AA \ apart in \nyo, and $f$-orbital overlap can induce CEF dispersion as is commonly found in closely-spaced non-dilute Ln-ion materials \cite{gaudet_gaulin, fulde_loewenhaupt}. This effect is not visible in the $E_i = 150$ meV data as the instrumental resolution at that energy transfer exceeds the observed splitting of $\sim$ 2.1 meV for all CEF excitations (Figure \ref{fig:F1}). A strong exchange-induced splitting in the powder averaged mode energies is consistent with the calculated susceptibilities for the CEF fits in Figure \ref{fig:F2} only coinciding with raw \nyo data if the previously determined $\theta_{CW} = -10.3$ K \cite{paper1NYO} is implemented. For reference, compared with the other Yb-based materials, $|\theta_{CW}|$ is more than two times larger than YbMgGaO$_4$ ( $|\theta_{CW}| = 4$ K) \cite{li_zhang, li_zhang2, li_zhang3, li_zhang4, paddison_mourigal, shen_zhao, xu_li} and an order of magnitude larger than Yb$_2$Ti$_2$O$_7$  ($|\theta_{CW}| = 0.4$ K) \cite{gaudet_gaulin, Bramwell_1999, hodges, ortenzio_luke}. 

At lower energies, the zero-field INS spectrum shows a diffuse continuum of excitations roughly centered about the $\bm{Q} =(1/3, 1/3, 0)$ position and emblematic of excitations within a highly frustrated, quantum disordered state \cite{paper1NYO, ding_tsirlin}. This continuum has a bandwidth of approximately 1 meV, and it sharply contrasts the expectations of coherent, dispersive modes from spin wave analysis of a low temperature $120^\circ$ ordered state (Figure \ref{fig:LSWT1}). Spectral weight at the K-point $\bm{Q} =(1/3, 1/3, 0)$ is present down to the lowest energies resolved $E=0.1$ meV, however, a peak appears in the spectral weight of the continuum at $E=0.25$ meV.  This suggests that part of the spectrum is gapped, for instance through damped magnons mixed within the continuum \cite{ferrari2019dynamical}, or that this feature is inherent to the structure factor of the continuum itself.

With increasing field, the ground state degeneracy is lifted and \nyo enters the classically-understood, ordered regime. By 2.75 T, evidence for the ordered \uud phase appears as the spin liquid phase recedes, and the INS data develop a powder-averaged, flat band that increases in energy with increasing field. This tracks spin wave predictions for the canted \uud phase in Figure \ref{fig:LSWT1}.  While this band is pushed up, another powder-averaged low energy band appears pushed downward toward zero energy near the previously reported high field phase boundary \cite{paper1NYO}. The mode softening at this upper field boundary may indicate the appearance of another, nearby magnetically ordered state prior to entering the quantum paramagnetic regime.  Previous NMR studies have suggested such a boundary exists \cite{ranjith_baenitz}, and numerical cluster mean-field methods predict a first order boundary into a neighboring V-state near 70\% of the saturation field \cite{yamamoto_danshita}.  The interpolated mode softening occurs near $H\approx 9.5$ T which is consistent with the $H_{sat}\approx 14$ T for NaYbO$_2$ \cite{paper1NYO, ding_tsirlin, ranjith_baenitz}.    

Further supporting the presence of a first-order high-field phase boundary in NaYbO$_2$, neutron diffraction data plotted in Fig. 6 demonstrate an irreversibility in spin correlations that arises upon crossing into the ordered state from the low- versus the high-field boundaries. The nature of this phase boundary merits further investigation, and it does not appear in the 2D classical phase diagram plotted in Fig. 7.  This suggests it likely derives from interplane coupling terms, consistent with predictions of numerical mean field models \cite{yamamoto_danshita}.  While the collinear \uud state emerges upon crossing into the ordered regime from below, the appearance of the previously absent $\bm{Q}=(1/3, 1/3, 2)$ peak upon entering the ordered regime from above implies a more noncollinear phase, consistent with a V-phase although single crystal measurements are required to fully unravel the nature of this hysteretic transition.         
 
Another feature that merits further study with single crystals is the appearance of resolvable magnetic scattering above the single-magnon cutoff in the \uud state.  The ability to resolve a broad band of scattering near 1.5 meV at $E^*\approx 3J$ suggests substantial magnon-magnon interactions and spectral weight transfered into the multimagnon sector of the INS spectrum \cite{mourigal_zhitomirsky, kamiya_ma}.  This would be consistent with presence of strong quantum fluctuations responsible for the reduction of the ordered moment and for transferring scattering weight into longitudinal spin fluctuations.  Earlier single crystal measurements of the \uud state of Ba$_3$CoSb$_2$O$_9$ failed to identify similar scattering \cite{kamiya_ma}, but quantum fluctuations are likely substantially enhanced in NaYbO$_2$ due to its strong interlayer frustration.   
 
In fact, the $ABC$ stacking sequence of layers in \nyo is frustrated relative to the $AAA$ stacking in Ba$_3$CoSb$_2$O$_9$ and related materials \cite{kamiya_ma, ma_matsuda, rawl_ma, cui_yu, kojima_avdeev, koutroulakis_brown, shirata_kindo}. Unfrustrated bonds between layers are generated when ions reside directly above each other ($AAA$), while frustration with three equivalent bonds forms when ions projected onto neighboring sheets reside at the centers of the triangular lattice. For instance, simulations of Ising moments on triangular lattice antiferromagnets with $ABC$ and $ABAB$ stacking have indeed shown that the frustrating interlayer interaction pushes magnetic ordering lower in temperature in comparison to $AAA$ stacking \cite{liu_chalker}. Remarkably, due to its structure-type promoting strong in-plane and inter-plane frustration, \nyo is joined by a host of $ALnQ_2$ compounds ($A =$ alkali ion, $Ln =$ Lanthanide ion, $Q =$ chalcogenide anion) to emerge in recent literature \cite{liu_zhangAMX2, paper1NYO, ding_tsirlin, ranjith_baenitz, xing_sefat, xing_sefat2, xing_sefat3,ranjith_baenitz2, sarkar_klauss} with seemingly quantum disordered magnetic ground states. For instance, both NaYbS$_2$ \cite{sarkar_klauss, sichelschmidt_doert, baenitz_doert} and NaYbSe$_2$ \cite{ranjith_baenitz2} have been proposed to display quantum spin liquid ground states. Other $Ln$ species such as Ce \cite{xing_sefat2} and Er \cite{xing_sefat} have also been investigated within this lattice type as potential hosts of quantum disordered magnetic states.  This suggests a rich realm of materials for exploring the XXZ Hamiltonian in equation \ref{Hamiltonian_J_D_B}--one in which in the $ALnQ_2$ compounds of varying local character can be used to explore the potential unconventional states predicted at this frontier.

\section{VI. Conclusions}
\nyo stands as a chemically-ideal quantum spin liquid candidate stabilized on an equilateral triangular lattice. The intramultiplet structure and CEF ground state doublet of Yb$^{3+}$ ions in \nyo comprised of mixed $|\pm 1/2 \rangle$, $|\pm 5/2 \rangle$, and $|\pm 7/2 \rangle$ states was identified via high-energy inelastic neutron scattering measurements. A subtle peak splitting is identified in the ground state's excitation spectrum that cannot be explained by trivial disorder and is, instead, consistent with weak CEF dispersion due to Yb exchange in this compound. The evolution of the low-energy inelastic spectra of \nyo is also reported. While low-temperature, zero-field powder-averaged data reveal a weighted continuum of excitations without magnetic order or spin freezing, upon increasing field, the spectra evolve into that predicted for an \uud magnetically-ordered phase. At the high-field phase boundary of the ordered state, a field-hysteretic diffraction pattern appears and suggests a first-order phase boundary into a noncollinear state.  Indications of strong magnon-magnon interactions are shown through the observation of appreciable spectral weight in the multimagnon spectrum of the field-induced ordered state.  Future neutron scattering experiments on single crystal specimens of  NaYbO$_2$ and related compounds are highly desired to resolve the detailed phase boundaries and higher order spin dynamics of the field-stabilized order.

\begin{acknowledgments}
This work was supported by the US Department of Energy, Office of Basic Energy Sciences, Division of Materials Sciences and Engineering under award DE-SC0017752 (S.D.W. and M.B.). M.B. acknowledges partial support by the National Science Foundation Graduate Research Fellowship Program under grant no. 1650114. Work by L.B. and C.L. was supported by the DOE, Office of Science, Basic Energy Sciences under award no. DE-FG02-08ER46524. P.M.S. acknowledges financial support
from the California Nanosystems Institute at UCSB, through the Elings Fellowship.  Identification of commercial equipment does not imply recommendation or endorsement by NIST.  The research that used the ARCS and CNCS facilities at the Spallation Neutron Source, as well as support for DP and AB during the experiment, is funded by the Department of Energy, Office of Science, Scientific User Facilities Division, operated by the Oak Ridge National Laboratory.
\end{acknowledgments}

\bibliography{NYO_paper2_bib_v9}

\end{document}